\documentclass{aa}

\providecommand{\orcid}[1]{}

\usepackage{graphicx}
\usepackage{amsmath,bm}
\usepackage{txfonts}
\usepackage{xcolor}
\usepackage{url}
\usepackage[colorlinks=true,linkcolor=blue,citecolor=blue,urlcolor=blue]{hyperref}
\usepackage{orcidlink}

\newcommand{\dmunits}{\mathrm{pc\,cm^{-3}}}

\begin{document}

\title{First fast radio burst search campaign at the Argentine Institute of Radio Astronomy}

\titlerunning{First FRB search campaign at the IAR}
\authorrunning{C. O. Lousto et al.}

\author{Carlos O. Lousto\inst{1}\orcidlink{0000-0002-6400-9640}\corrauth{colsma@rit.edu}
   \and Harsh Prajapati\inst{2}
   \and Ezequiel Zubieta\inst{3}\orcidlink{0009-0009-5593-367X}
   \and Susana B. Araujo Furlan\inst{4,5}\orcidlink{0000-0003-4027-4826}
   \and Guillermo Gancio\inst{3}\orcidlink{0000-0003-1282-3031}
   \and Joaqu\'in Pelle\inst{6}\orcidlink{0000-0001-5820-8208}
   \and Gustavo E. Romero\inst{3}\orcidlink{0000-0002-5260-1807}
   \and Federico Garc\'ia\inst{3}\orcidlink{0000-0001-9072-4069}
   \and Santiago del Palacio\inst{3,7}\orcidlink{0000-0002-5761-2417}
   \and Nathan Cahill\inst{8}\orcidlink{0000-0002-6164-3291}
}

\institute{Center for Computational Relativity and Gravitation, Rochester Institute of Technology, Rochester, NY 14623, USA
      \and Center for Imaging Science, Rochester Institute of Technology, Rochester, NY 14623, USA
      \and Instituto Argentino de Radioastronom\'ia (CCT La Plata, CONICET; CICPBA; UNLP), C.C.5, (1894) Villa Elisa, Buenos Aires, Argentina
      \and Facultad de Matem\'atica, Astronom\'ia, F\'isica y Computaci\'on, UNC, Av. Medina Allende s/n, Ciudad Universitaria, X5000HUA C\'ordoba, Argentina
      \and Instituto de Astronom\'ia Te\'orica y Experimental, CONICET-UNC, Laprida 854, X5000BGR C\'ordoba, Argentina
      \and Max Planck Institute for Gravitational Physics (Albert Einstein Institute), Am M\"uhlenberg 1, 14476 Potsdam, Germany
      \and Department of Space, Earth and Environment, Chalmers University of Technology, SE-412 96 Gothenburg, Sweden
      \and School of Mathematics and Statistics, Rochester Institute of Technology, Rochester, NY 14623, USA
}

\date{Received XXX; accepted XXX}

\abstract
{Fast radio bursts (FRBs) are intense millisecond-duration radio transients of extragalactic origin whose physical nature remains under active investigation, and which also serve as probes of the intergalactic medium.}
{We report on the first FRB search campaign carried out at the Argentine Institute of Radio Astronomy (IAR) between December 2024 and March 2026, targeting nearby galaxy superclusters in the southern sky.}
{We observed fields in the Ophiuchus, Shapley, and Sculptor/Phoenix supercluster regions with one of the two 30~m antennas of the IAR, using a ROACH-based backend with 400~MHz of bandwidth centred at 1400~MHz and a time resolution of 41--82~$\mu$s, for a total net observing time of 212~h. The data were searched for dispersed single pulses with \texttt{PRESTO} in the dispersion measure range $100 \leq \mathrm{DM} \leq 500~\dmunits$, and candidates were classified with the FETCH machine learning classifier. The pipeline was validated on archival Parkes data containing known FRBs and on synthetic bursts injected into IAR observations.}
{One FRB candidate, FRB~20251018, was identified on 18 October 2025 in an observation pointed towards the galaxy cluster A2870, in the Phoenix supercluster, with a dispersion measure of $243~\dmunits$, a signal-to-noise ratio of 8.2, and a FETCH probability of $p=0.99$. To the best of our knowledge, this would be the first FRB detected from South America. A set of more marginal candidates is also presented.}
{These results demonstrate the capability of the IAR antennas to detect FRBs and support the continuation and extension of the monitoring campaign, including coincident dual-antenna observations and cross-matches with gravitational-wave events and electromagnetic transients.}


\maketitle
\nolinenumbers

\section{Introduction}\label{sec:Intro}

The Argentine Institute of Radio Astronomy (IAR), located at latitude $-34\degr 51\arcmin 57\farcs35$ and longitude $-58\degr 08\arcmin 25\farcs04$, operates two 30~m single-dish antennas, aligned along the North--South direction and separated by $120$~m. They cover a declination range of $-90\degr < \delta < -10\degr$ and an hour-angle range of two hours east/west, $-2\,\mathrm{h} < t < 2\,\mathrm{h}$. Observations are conducted using both radio telescopes, named after Carlos M. Varsavsky and Esteban Bajaja (hereafter A1 and A2), in dual polarisation (with both circular components combined) centred at 1400~MHz (L~band).

In \citet{Gancio:2019frj} we provided a thorough description of the initial Ettus configuration of the front ends of A1 and A2. Furthermore, the analysis of the radio frequency interference (RFI) environment presented in \citet{Gancio:2019frj} revealed that the radio band from 1~GHz to 2~GHz has a low level of RFI activity, which is adequate for radio astronomy, despite the fact that the IAR is not located in an RFI-quiet zone. In addition to the Ettus boards, in mid-2022 we added a parallel digitiser system based on Reconfigurable Open Architecture Computing Hardware (ROACH) boards \citep{hickish2016casper}. The ROACH boards are configured to observe with both antennas in dual circular polarisation over a bandwidth of $400~\mathrm{MHz}$. The new ROACH-1 backends\footnote{\url{https://casper.astro.berkeley.edu/wiki/ROACH}}$^,$\footnote{\url{https://digicom.org/roach-board.html}}, installed on both A1 and A2 with a bandwidth of 400~MHz and integration times down to 41 or 82~$\mu$s, provide a theoretical improvement by a factor of up to 2.5 in the signal-to-noise ratio (S/N) with respect to the Ettus boards with 56~MHz of bandwidth, thus notably improving the quality of the observations and the prospects of detecting new FRBs.

Our observational schedule focuses on high-cadence observations, reaching daily cadence for some pulsars \citep{2025A&A...698A..72Z,2025A&A...694A.124Z, 2024A&A...689A.191Z,2023MNRAS.521.4504Z} and magnetars \citep{2026A&A...710A..13A}, with individual observations as long as 3.66 hours per day.

In this work we use the data from the A2 ROACH receiver to present the first blind FRB search campaign carried out at the IAR, covering observations taken between December 2024 and March 2026.

\section{Fast radio bursts: a brief review}\label{sec:FRBs}

Fast radio bursts (FRBs) are intense millisecond-duration radio pulses of extragalactic origin. Despite their relatively recent discovery, their study has advanced rapidly during the last decade, and they are widely recognised not only for their intrinsic astrophysical interest but also for their potential as cosmic probes.

The first FRB was discovered in 2007, upon reanalysing data from a survey of pulsars in the Small Magellanic Cloud (SMC) \citep{lorimer2007bright}. It was a single burst whose dispersion measure (DM) of $\sim 375~\dmunits$ greatly exceeded that expected from the Milky Way or the SMC. This emblematic event (named the ``Lorimer burst'') therefore seemed to have an extragalactic origin. However, even its celestial origin remained in doubt due to the absence of additional detections during the following years, in addition to the appearance of similar signals, known as \textit{perytons}, which turned out to be terrestrial interference caused by microwave ovens, clustered around dining time \citep{petroff2015identifying}. In 2013, the detection of four new FRBs, outside the Galactic plane, with DMs up to $1100~\dmunits$---much greater than that of any peryton, and crucially not clustered around dining time---established that FRBs were genuine astrophysical phenomena \citep{thornton2013population}. Since then, thousands of events have been detected \citep{amiri2021first}, most of them with DMs in the range of $200$--$1200~\dmunits$ and typical distances from the order of $100~\mathrm{Mpc}$ up to a few $\mathrm{Gpc}$. The event rate of FRBs is estimated to be around $10^4~\mathrm{sky}^{-1}\,\mathrm{day}^{-1}$ above a threshold fluence of $\sim 3~\mathrm{Jy}\,\mathrm{ms}$ at $1.4~\mathrm{GHz}$ \citep{thornton2013population}, and their intrinsic volumetric rate density is $\sim 10^7~\mathrm{Gpc}^{-3}\,\mathrm{yr}^{-1}$ for luminosities above $10^{38}\,\mathrm{erg}\,\mathrm{s}^{-1}$ \citep{ravi2019prevalence, lu2020unified}.

Most FRBs have been detected at frequencies between $400~\mathrm{MHz}$ and $1.4~\mathrm{GHz}$, but there are confirmed events from $\sim 100~\mathrm{MHz}$ \citep{pleunis2021lofar} up to $8~\mathrm{GHz}$ \citep{gajjar2018highest}. The typical durations are of the order of milliseconds and, in some cases, substructures are identified on scales down to tens of microseconds \citep{farah2018frb}. Most events consist of a single pulse, but bursts with multiple components are common as well \citep{zhou2022fast, champion2016five}. Some FRB pulses are asymmetric, with a long decaying tail, as expected from scattering in a cold plasma. Some also show frequency down-drifting of subpulses \citep{zhou2022fast, hessels2019frb}. The peak flux densities vary between $\sim 0.1$ and tens of Jy, with typical fluences around $1$--$100~\mathrm{Jy}\,\mathrm{ms}$ \citep{bannister2017detection, niu2021crafts}. Their distribution on the sky is approximately isotropic, consistent with their extragalactic origin \citep{bhandari2018survey}.

A fraction of FRBs are observed to repeat, with waiting times between bursts ranging from milliseconds up to weeks \citep{spitler2016repeating, chime2020periodic}. The most emblematic case is FRB~121102, which enabled the first accurate localisation and identification of a host galaxy, at $z \sim 0.19$ \citep{chatterjee2017direct}, consolidating the hypothesis of a cosmological origin. From reliable localisations, a correlation between the total DM and the redshift $z$ has been observationally established \citep{macquart2020census}. In principle, all apparently one-off FRBs could be repeaters with very long waiting times. However, some differences are observed between the populations of repeaters and apparent non-repeaters \citep{amiri2021first, pleunis2021fast}: while repeaters tend to have broader widths, complex morphologies, and narrower spectra, single bursts have simpler pulse shapes and narrower widths.

Many FRBs exhibit a high degree of linear polarisation (frequently $> 80\%$), some with rapid variations of the polarisation angle during the event \citep{xu2022fast, michilli2018extreme, luo2020diverse, day2020high}. The Faraday rotation measure (RM), which depends on the magnetic field along the line of sight, varies significantly among events. While some FRBs present values compatible with zero, others---like FRB~121102---have extreme RMs of the order of $10^5~\mathrm{rad}\,\mathrm{m}^{-2}$ \citep{michilli2018extreme}, which suggests a very dynamic, highly magnetised local environment.

In most cases, no counterpart at other wavelengths has been detected, despite significant observational efforts \citep{zhang2024multiwavelength}. The only robust exception is the Galactic FRB associated with the magnetar SGR~1935+2154, observed in 2020 in coincidence with an X-ray burst \citep{bochenek2020fast, mereghetti2020integral}. This association established for the first time---and the only one so far---the nature of the astrophysical source of an FRB.

The combination of short durations, high fluences, and cosmological distances implies isotropic-equivalent energies up to $10^{43}$~erg and brightness temperatures exceeding $10^{35}$~K \citep{thornton2013population}. Such values can only be explained by coherent emission mechanisms under extreme physical conditions \citep{katz2016fast}: compact emission regions (from tens to hundreds of kilometres), high magnetisations, and dense plasmas.

Several candidates have been proposed for the central engine of FRBs \citep{zhang2023physics}. The most observationally supported are young and highly magnetised neutron stars, i.e., magnetars. In these objects, internal perturbations or external interactions can trigger sudden releases of magnetic energy, capable of driving relativistic shocks or magnetospheric excitations that produce coherent radio emission \citep{metzger2019fast, kumar2020frb, lyubarsky2020fast, vanthieghem2025fast, mahlmann2022electromagnetic}. The emission detected from SGR~1935+2154 corresponds to an energy of $\sim 10^{35}$~erg \citep{bochenek2020fast}, much smaller than that of extragalactic FRBs, but similar in morphology and temporal features. This suggests the same physical mechanism operating at different energy scales, and that at least a fraction of FRBs come from magnetars.

Finally, the observed DM--$z$ correlation enables the use of FRBs as probes of the intergalactic medium (IGM). The DM--$z$ correlation is approximately linear, as expected if most of the dispersion originates in the IGM. In particular, it has been demonstrated that the dispersion attributable to the IGM is consistent with that inferred from observations of the cosmic microwave background and primordial nucleosynthesis, solving the long-standing ``missing baryon problem'' and suggesting that the missing baryons reside in the IGM \citep{mcquinn2013locating, macquart2020census}. In addition, the low RMs observed in many distant FRBs impose an upper bound on the mean IGM magnetic field \citep{akahori2016fast}. With larger samples, FRBs could also be used to constrain dark energy \citep{walters2018future, kumar2019use}, the reionisation history \citep{beniamini2021exploring}, and the large-scale structure of the universe \citep{shirasaki2022probing, rafiei2021chime}. If strongly lensed FRBs are detected in the future, which is likely given the high event rate of FRBs, they could also be used to measure the Hubble constant and the curvature of the universe, and to constrain dark matter \citep{li2018strongly}. As more events are observed, FRBs could help not only to clarify the extreme physics of their central engines, but also to address important questions about the composition and evolution of the universe.


\section{Techniques}\label{sec:Techniques}

With thousands of hours of good-quality pulsar observations, it is interesting to investigate whether the data also contain FRB signals \citep{Petroff:2019tty}. In particular, machine learning techniques have been developed to perform massive searches for FRBs \citep{Zhang:2018jux} and to classify them \citep{Connor:2018wfr,Wagstaff:2016pdw}, using both supervised \citep{Luo:2022smj} and unsupervised methods \citep{Chen:2021jpq,Zhu-Ge:2022nkz}. A practical implementation for fast transient classification \citep{2020MNRAS.497.1661A} is FETCH\footnote{\url{https://github.com/devanshkv/fetch}}. Another useful tool is the synthetic FRB generator \texttt{FRB-faker}\footnote{\url{https://gitlab.com/houben.ljm/frb-faker}}, which can be used to train FRB search and classification algorithms. A living theoretical catalogue of FRBs, with a review of the numerous existing theories proposed to model them, is presented in \citet{Platts:2018hiy}; see also the list of theoretical scenarios at \url{https://frbtheorycat.org/index.php/Main_Page} and the list of observed FRBs at \url{https://www.frbcat.org}. Another FRB catalogue is reported in \citet{universe9070330}\footnote{\url{https://www.mdpi.com/2218-1997/9/7/330}}. Further work relating FRBs and machine learning techniques can be found in \citet{Connor:2018wfr,Wagstaff:2016pdw,Zhang:2018jux}. 

The Petabyte Project maintains a comprehensive list of FRB software\footnote{\url{https://github.com/thepetabyteproject/frbsoft}}, which includes: data read and write packages (\texttt{your}), single-pulse search (\texttt{PRESTO}), single-pulse injection (\texttt{FRB-faker}), machine learning classifiers (\texttt{FETCH}), population synthesis of FRBs, catalogues and tools for catalogues, burst analysis software, visualisations, RFI mitigation (\texttt{RFIClean}), and containers. The codes given in parentheses are those with which we have hands-on experience. New FRB findings can also be cross-checked for coincidences with gravitational-wave detections, as recently attempted in \citet{LIGOScientific:2022jpr}.

Our single-pulse search pipeline proceeds as follows. The dedispersion plan is generated with \texttt{DDplan.py}, and the filterbank data are dedispersed with \texttt{prepsubband} (\texttt{PRESTO}) using 32 subbands over a grid of trial DMs covering the chosen search range $100 \leq \mathrm{DM} \leq 500~\dmunits$ in steps of $1~\dmunits$, with the output time series downsampled by a factor of eight (to 0.33 and 0.66~ms for the observations sampled at 40.96 and 81.92~$\mu$s, respectively). The lower bound safely excludes Galactic sources and low-DM local interference, since the expected Milky Way disc contribution along our high-latitude target sightlines is only $\sim 20$--$30~\dmunits$ (see Sect.~\ref{sec:Discussion}), while the upper bound---which encompasses the DMs expected for sources in or moderately behind the targeted superclusters, and the peak of the observed FRB DM distribution \citep{amiri2021first}---was chosen to limit the computational cost of this first processing of the data. The resulting time series are searched with \texttt{single\_pulse\_search.py} with a detection threshold of $8\sigma$ and boxcar widths of up to 45 samples (\texttt{max\_downfact}), i.e. up to $\simeq 29$~ms at the downsampled resolution. The surviving candidates are converted into HDF5 cutouts with \texttt{your\_candmaker.py} \citep[from the \texttt{your} package;][]{Aggarwal:2020ulu} and classified with the FETCH deep-learning classifier (model~A), retaining candidates with $p>0.5$ for visual inspection, where $p$ is the probability score assigned by the network to a candidate being a genuine astrophysical burst rather than RFI \citep{2020MNRAS.497.1661A}. This last step provides the bulk of the data reduction: a typical observation yields several thousand single-pulse triggers above the $8\sigma$ threshold---the overwhelming majority of them low-level RFI---of which FETCH retains of order half a dozen with $p>0.5$, a number small enough for every surviving candidate to be inspected individually by eye. The validation of this pipeline, on archival Parkes data containing seven known FRBs and on synthetic bursts injected into IAR observations, is described in detail in Appendix~\ref{app:Tables}: all seven archival FRBs were recovered with the correct arrival times and dispersion measures, and all six synthetic bursts (with $10 \leq \mathrm{S/N} \leq 45$ and $120 \leq \mathrm{DM} \leq 874~\dmunits$), including one administered as a blind test, were recovered by the pipeline.

\section{Observations at IAR}\label{sec:Observations}

There are essentially two strategies one can follow to observe FRBs with the IAR antennas. The first is to observe regions of the sky that are expected to produce FRBs, such as nearby galaxy clusters. The second is to track known FRBs in order to look for repeaters. Given the rates of up to 10\,000~FRBs sky$^{-1}$ day$^{-1}$ reported in Table~3 of \citet{Petroff:2019tty}, together with the large amount of observing time available at our antennas, we can make a first estimate of the rate of FRBs potentially detectable by the IAR as $\dot{N}_\mathrm{IAR} \sim \dot{N}_\mathrm{FRB} \times \Delta\Omega/(4\pi)$, with $\Delta\Omega \approx \pi \theta_\mathrm{FWHM}^2/4$. This first guess suggests that a detection is likely after $\sim 1000$~h of accumulated observations.

The biggest challenge in the search for fast radio bursts is that we do not know when or where the next one will occur. The radio sky must therefore be observed continuously until one of these events is captured. Since thousands of FRBs are thought to occur per day, serendipitous detections are feasible. This implies that large amounts of raw radio-astronomical data must be accumulated and analysed. Observations with a bandwidth of 400~MHz, sampled every 82~$\mu$s in 512 frequency channels of 0.78~MHz, generate a typical data volume of 85~GB per hour of observation, corresponding to up to 2~TB per day and 0.75~PB per year. Given that the IAR observatory can track sources for up to 3.66~h per day, any single observation may potentially contain an FRB signal, motivating the application of machine learning algorithms to search for such patterns in the accumulated raw data. As an illustration of the potential of such archival searches, \citet{Zhang:2019xzf} reported the discovery of a new FRB (FRB~010312) in the original archival data set of the first detection (FRB~010724), almost two decades later.

In our search for FRBs in the southern sky with A2, which has a primary beam of $30\arcmin$ (FWHM), we avoid the Galactic Centre (RA 17h 45m 40.05s, Dec $-29\degr 00\arcmin 28\farcs1$) and the Galactic plane, focusing instead on nearby superclusters in the southern hemisphere that are also accessible within the observatory schedule. All coordinates quoted in this paper are in the J2000 epoch.

All fields were observed with the same standard configuration of the A2 ROACH backend: dual circular polarisation (summed), with a bandwidth of 400~MHz divided into 512 channels of 0.78125~MHz, spanning 1200.8--1600.0~MHz (centred at 1400.4~MHz), and recorded as 32-bit SIGPROC filterbank files. The Ophiuchus and Shapley fields, observed between December 2024 and January 2025, were sampled at 40.96~$\mu$s, whereas the longer Sculptor and Phoenix campaigns were carried out at 81.92~$\mu$s to halve the data volume; both sampling times vastly oversample the expected millisecond burst widths, given that the intra-channel dispersion smearing at $\mathrm{DM} \simeq 250~\dmunits$ already amounts to $\sim$0.6~ms at the centre of the band. Individual pointings typically lasted between $\sim$0.8 and $\sim$2~h (e.g., 2920~s and 4632~s for representative Ophiuchus and Shapley observations, and $\sim$7110--7130~s for the A4013 and A2870 fields), within the maximum tracking time of 3.66~h.

\paragraph{Ophiuchus supercluster} (RA 17h 10m 00s, Dec $-22\degr 00\arcmin 00\arcsec$):
The Ophiuchus supercluster is a nearby galaxy supercluster in the constellation Ophiuchus, which forms the far wall of the Ophiuchus Void. It is centred on the cD cluster (Abell class type~I) Ophiuchus Cluster and includes at least two additional galaxy clusters, four galaxy groups, and several field galaxies. Our observations focus on the central region of the supercluster. Notably, the Ophiuchus Cluster hosts a sharply truncated cool core \citep{2016MNRAS.460.2752W} and a giant cavity, $\sim 0.5$~Mpc across, filled with low-frequency radio emission that has been interpreted as the fossil of one of the most energetic active-galactic-nucleus outbursts known, powered by the supermassive black hole of the central galaxy \citep{2020ApJ...891....1G}. A summary of the observations analysed is given in Table~\ref{tab:sc1710_2200}.

\begin{table}
    \centering
    \caption{Observation summary for the Ophiuchus supercluster field (Dec $-22\degr 00\arcmin 00\arcsec$, RA 17h 10m 00s).}
    \begin{tabular}{lccc}
        \hline\hline
        Month & \#Obs & Total (s) & Total (h) \\
        \hline
        2024-12 & 2 & 5896.28  & 1.6378 \\
        2025-01 & 3 & 8798.17  & 2.4439 \\
        \hline
        TOTAL   & 5 & 14694.45 & 4.0817 \\
        \hline
    \end{tabular}
    \label{tab:sc1710_2200}
\end{table}

\paragraph{Shapley supercluster} (RA 13h 25m 00s, Dec $-30\degr 00\arcmin 00\arcsec$):
Identified as a single, exceptionally massive concentration of clusters in 1989 \citep{1989Natur.338..562S, 1989Natur.342..251R}, it is named after Harlow Shapley, who first noted an excess of galaxies in this region of the sky in the 1930s. The Shapley supercluster is a massive structure that has been the subject of numerous studies; although it is not the largest known supercluster, it is certainly one of the densest. It contains at least twenty rich galaxy clusters, among thousands of galaxy groups, including three of the richest galaxy clusters known: A3558, A3559, and A3560. We focus on a fairly central location near A3556 and A1736 in order to include several of these galaxy clusters within the primary beam of the radio telescope. A summary of the observations analysed is given in Table~\ref{tab:sc1325_3000}.

\begin{table}
    \centering
    \caption{Observation summary for the Shapley supercluster field (Dec $-30\degr 00\arcmin 00\arcsec$, RA 13h 25m 00s).}
    \begin{tabular}{lccc}
        \hline\hline
        Month & \#Obs & Total (s) & Total (h) \\
        \hline
        2024-12 & 10 & 53418.03 & 14.8383 \\
        2025-01 & 2  & 9981.42  & 2.7726 \\
        \hline
        TOTAL   & 12 & 63399.45 & 17.6109 \\
        \hline
    \end{tabular}
    \label{tab:sc1325_3000}
\end{table}

\paragraph{Sculptor region} (multiple pointings; Sculptor Galaxy: RA 23h 13m 14.02s, Dec $-28\degr 21\arcmin 26\farcs94$):
Two superclusters in the Sculptor and Phoenix regions of the sky mark the position of a very long wall composed of thousands of galaxy groups stretching over nearly 300~Mpc, probably the longest of the nearby galaxy walls. The Sculptor superclusters are particularly important because they lie within this major wall of galaxies. We focus on a central region around the rich Abell cluster A4013 \citep[RA 23h 31.9m, Dec $-35\degr 16\arcmin$, $z \simeq 0.049$, i.e., $d \approx 210$~Mpc for $H_0 = 70$~km\,s$^{-1}$\,Mpc$^{-1}$;][]{1989ApJS...70....1A, 1999ApJS..125...35S, valotto2004faint}. A summary of the observations analysed is given in Table~\ref{tab:sc2331_3516}.

\begin{table}
    \centering
    \caption{Observation summary for the Sculptor supercluster field around A4013 (Dec $-35\degr 16\arcmin 00\arcsec$, RA 23h 31m 54s).}
    \begin{tabular}{lccc}
        \hline\hline
        Month & \#Obs & Total (s) & Total (h) \\
        \hline
        2025-04 & 3  & 17865.19  & 4.9625 \\
        2025-05 & 2  & 10262.32  & 2.8506 \\
        2025-07 & 2  & 19323.01  & 5.3675 \\
        2025-08 & 7  & 45731.58  & 12.7032 \\
        2025-09 & 6  & 42826.84  & 11.8963 \\
        2025-10 & 11 & 80028.13  & 22.2300 \\
        2026-01 & 5 & 19016.32 & 5.2823 \\
        2026-02 & 7 & 27432.12 & 7.6200 \\
        2026-03 & 10 & 36520.24 & 10.1445\\
        \hline
        TOTAL & 53 & 299005.74 & 83.0571\\
        \hline
    \end{tabular}
    \label{tab:sc2331_3516}
\end{table}

The nearer of the two superclusters lies mainly within the constellation Phoenix and is therefore perhaps better referred to as the Phoenix supercluster. The six clusters in the Phoenix supercluster are all of richness class zero, meaning that none of them is particularly prominent. Our observations focus on a region centred around A2870 \citep[RA 01h 07.7m, Dec $-46\degr 55\arcmin$, $z=0.0225$, i.e., $d \approx 96$~Mpc;][]{1989ApJS...70....1A, 1999ApJS..125...35S}. A summary of the observations analysed is given in Table~\ref{tab:sc0107_4655}.

\begin{table}
    \centering
    \caption{Observation summary for the Phoenix supercluster field around A2870 (Dec $-46\degr 55\arcmin 00\arcsec$, RA 01h 07m 42s).}
    \begin{tabular}{lccc}
        \hline\hline
        Month & \#Obs & Total (s) & Total (h) \\
        \hline
        2025-04 & 8  & 39391.66  & 10.9421 \\
        2025-05 & 9  & 56144.47  & 15.5956 \\
        2025-06 & 9  & 38092.31  & 10.5811 \\
        2025-07 & 14 & 45234.42  & 12.5651 \\
        2025-08 & 10 & 58225.58  & 16.1737 \\
        2025-09 & 6  & 42758.98  & 11.8774 \\
        2025-10 & 10 & 62156.69  & 17.2657 \\
        2026-01 & 2 & 13485.02 & 3.7458\\
        2026-02 & 3 & 16531.78 & 4.5921\\
        2026-03 & 2 &14274.41 & 3.9651\\
\hline
        TOTAL & 73 & 386295.31 & 107.3041\\
        \hline
    \end{tabular}
    \label{tab:sc0107_4655}
\end{table}

In processing these observations with \texttt{rfifind}/\texttt{PRESTO}, we applied an aggressive mask to filter out any potential source of local interference, thus reducing the effective bandwidth from the original 400~MHz to a net 264~MHz. Specifically, of the 512 channels into which the 400~MHz bandwidth is divided, we zapped channels 0:30, 63:71, 75:146, 170:175, 220:234, 273, 362:365, 382:385, 428:438, 441:444, 464, 476:490, and 511. The masked spectrum thus amounts to 135.94~MHz, and the clean (untouched) spectrum to 264.06~MHz. The single largest contiguous block of clean spectrum lies between channels 274 and 361, spanning 68.75~MHz of bandwidth. This static mask is applied identically to all four fields, and is supplemented by the per-observation flagging produced by \texttt{rfifind}, which removes a further, variable fraction of the band; in the observation in which FRB~20251018 was detected the two together leave $\simeq 170$~MHz of live bandwidth, spanning 1315--1520~MHz, and it is this effective value that we adopt for the flux-density estimate in Sect.~\ref{sec:Discussion}. Although this aggressive masking sacrifices $\sim$34\% of the nominal bandwidth---and hence $\sim$19\% in radiometer sensitivity---the interference is confined to stable, well-identified frequency ranges, so that the remaining 264~MHz of clean spectrum still exceeds the bandwidth of many successful single-dish FRB programmes; we therefore do not regard the RFI environment as a fundamental limitation for future campaigns at the IAR, although the mask will be re-evaluated periodically as the local spectrum occupancy evolves.

\section{Discussion and conclusions}\label{sec:Discussion}

Although we have not found any FRB candidates in our observations of the Ophiuchus and Shapley galaxy superclusters, this was expected given the relatively small number of hours accumulated for these targets. On the other hand, for the longer set of observations in the direction of the Sculptor region, we can report a candidate event.

In the observations of the cluster A2870 (declination $\delta=-46\degr 55\arcmin 00\arcsec$, right ascension $\alpha=$ 01h 07m 42s), we identified a candidate that has been preliminarily reported in an Astronomer's Telegram by our group as FRB~20251018 \citep{2026ATel17580....1P}. To the best of our knowledge, this would be the first FRB detected from South America.

\begin{figure}
    \centering
    \includegraphics[width=\hsize]{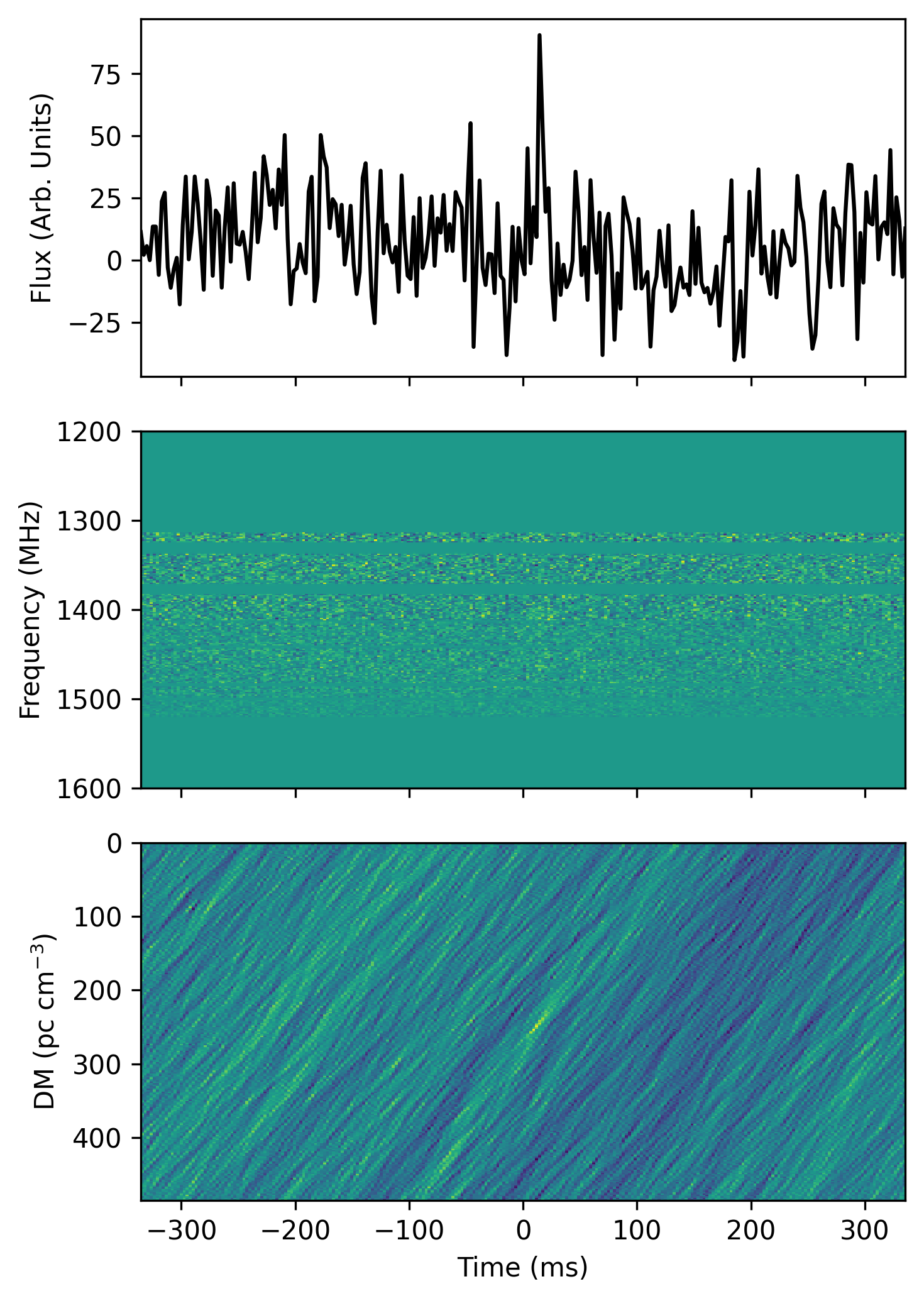}
    \caption{FRB candidate (FRB~20251018) at MJD 60966.21814523 in the direction of the cluster A2870 (Phoenix supercluster). Top: dedispersed pulse profile. Middle: dedispersed dynamic spectrum. Bottom: DM--time (\textit{bow-tie}) plane.}
    \label{fig:FRB1}
\end{figure}

This candidate appears in the observation of 18 October 2025, starting at MJD 60966.207662037035, at $t=905.7474970$~s after the start of the observation (i.e., at MJD 60966.21814523), with a dispersion measure $\mathrm{DM}=243\pm15~\dmunits$, an observed width of $W \approx 6$~ms (FWHM), with the burst envelope spanning $\approx 25$~ms at the base of the dedispersed profile shown in Fig.~\ref{fig:FRB1}, a signal-to-noise ratio $\mathrm{S/N}=8.20$, and a FETCH-assigned probability of $p=0.99$. The candidate was identified with \texttt{PRESTO} and validated by the FETCH machine learning classifier; based on the radiometer equation ---assuming a system temperature $T_\mathrm{sys} = 50\pm8$~K and an antenna gain $G \simeq 0.077$~K\,Jy$^{-1}$ for A2 (corresponding to an aperture efficiency of $\simeq 0.3$), i.e., a system equivalent flux density $\mathrm{SEFD} \simeq 650$~Jy \citep{Gancio:2019frj, 2026A&A...710A..13A}--- and the observed parameters, we estimated a band-averaged peak flux density of $\approx 3.0\pm0.5$~Jy and a fluence of $\mathcal{F} \approx 27\pm5$~Jy\,ms, adopting the boxcar-equivalent width $W_\mathrm{eq} = \mathcal{F}/S_\mathrm{peak} \approx 9$~ms measured from the dedispersed profile and the $\simeq 170$~MHz of bandwidth that survived flagging in this observation. The value of $3.9\pm0.9$~Jy quoted in the discovery report \citep{2026ATel17580....1P} follows from the same expression evaluated at the FWHM of the profile rather than at $W_\mathrm{eq}$. To confirm the astrophysical, broadband nature of the signal and to rule out narrowband RFI, we performed a hierarchical channel-zapping test, successively masking frequency channels in five stages down to the cleanest 170~MHz of the central band; the signal remained robust against these cuts \citep{2026ATel17580....1P}.
Figure~\ref{fig:FRB1} shows this FRB candidate. The quoted uncertainty on the DM is not the $1~\dmunits$ step of the search grid, which oversamples the time resolution, but the width of the signal-to-noise ratio versus trial-DM response: for a burst of this width and signal-to-noise ratio, detected over the $\simeq 170$~MHz that survived flagging, the response peaks at $\mathrm{DM} \simeq 240~\dmunits$ and is consistent with the adopted value over a range of $\simeq \pm 15~\dmunits$.

The expected Milky Way contribution to the DM along this high-Galactic-latitude sightline ($l=294\fdg8$, $b=-70\fdg0$) is small: the NE2001 \citep{cordes2002ne2001} and YMW16 \citep{yao2017ymw16} Galactic electron density models predict $\mathrm{DM_{MW}} \simeq 31$ and $20~\dmunits$, respectively, to which an additional Galactic halo contribution of $\sim 30$--$50~\dmunits$ may be added \citep{yamasaki2020halo}. The resulting extragalactic excess, $\mathrm{DM_{exc}} = \mathrm{DM_{obs}} - \mathrm{DM_{MW}} \simeq 170$--$220~\dmunits$, sets---via the Macquart relation \citep{macquart2020census}---a conservative redshift upper limit of $z \lesssim 0.2$ if the host contribution is negligible. This comfortably encompasses the targeted cluster A2870 at $z=0.0225$, for which the mean cosmic contribution is only $\sim 20~\dmunits$; an origin in A2870 would therefore require a combined host-galaxy and intracluster-medium contribution of $\sim 150$--$200~\dmunits$ in the source frame, larger than the typical values assumed for field FRBs but well within the broad observed distribution of host DMs, particularly for sources embedded in dense environments. Alternatively, the burst could originate from a background source at $z \lesssim 0.2$ seen through the cluster field.

With one detection in 212~h of observations with a $30\arcmin$ beam ($\Delta\Omega/4\pi \simeq 4.8\times10^{-6}$), our campaign implies an all-sky rate of order $\dot{N}_\mathrm{FRB} \sim 1/(212\,\mathrm{h} \times 4.8\times10^{-6}) \approx 2\times10^{4}~\mathrm{sky}^{-1}\,\mathrm{day}^{-1}$ above our sensitivity threshold, consistent within the (large, Poisson-dominated) uncertainties with the rates compiled in \citet{Petroff:2019tty}.

Five additional, more marginal candidates---with lower FETCH $p$-scores and/or diagnostic plots of lower quality than those of FRB~20251018, but surviving our visual inspection against RFI contamination---are listed in Table~\ref{Table:MoreFRB} in Appendix~\ref{app:marginal}, where the downselection criteria are described.

The results of this first campaign demonstrate that the IAR antennas, equipped with the ROACH backends, are capable of detecting FRBs, and they support the continuation and extension of the monitoring programme. The campaign will continue with extended observations of the Sculptor/Phoenix fields and, in particular, of the position of FRB~20251018, in order to search for possible repeat bursts from the same source. Monitoring of the A4013 and A2870 fields has in fact already continued through March 2026 (Tables~\ref{tab:sc2331_3516} and~\ref{tab:sc0107_4655}) without yielding a second firm detection: leaving aside the marginal candidates of Appendix~\ref{app:marginal}, we can put forward at most one detection per supercluster field, and no repeat burst from the position of FRB~20251018 has been identified so far. We also plan to carry out simultaneous observations with both IAR antennas, A1 and A2: the detection of a burst in coincidence in two independent receivers separated by 120~m would provide a decisive discriminant against local interference and instrumental artefacts, and would allow us to confirm candidates near the detection threshold, such as those listed in Appendix~\ref{app:marginal}.

The scientific return of the survey can be further enhanced by cross-matching our candidates with transients detected by other facilities. In particular, FRB candidates detected at the IAR can be searched for temporal and spatial coincidences with the gravitational-wave events reported by the LIGO--Virgo--KAGRA (LVK) Collaboration, along the lines of the joint CHIME/FRB--LVK analysis of \citet{LIGOScientific:2022jpr}; compact-binary mergers involving neutron stars are among the proposed FRB progenitor channels, and even null results provide useful constraints. Similarly, coincidences with electromagnetic counterparts at other wavelengths---most notably the high-energy bursts of magnetars, following the association of the Galactic FRB with SGR~1935+2154 \citep{bochenek2020fast, mereghetti2020integral}, but also with the broader population of X-ray, gamma-ray, and optical transients \citep{zhang2024multiwavelength}---can help to establish the nature of the sources. The flexible, high-cadence scheduling of the IAR antennas makes them well suited both to systematic monitoring and to rapid follow-up of such multimessenger alerts in the southern sky.

\section*{Data availability}
The filterbank observations analysed in this work, and the candidate cutouts of FRB~20251018 and of the marginal candidates listed in Appendix~\ref{app:marginal}, are available from the corresponding author upon reasonable request.

\begin{acknowledgements}
COL gratefully acknowledges support from NSF awards AST-2319326, PHY-2207920, and PHY-2513442. We are very grateful to the staff of the IAR for their continuous technical support during the execution of this work. S.B.A.F. and E.Z. are PhD candidates with CONICET fellowships. G.E.R. and F.G. are CONICET researchers. G.E.R. and F.G. acknowledge financial support from the State Agency for Research of the Spanish Ministry of Science and Innovation under grant PID2022-136828NB-C41/AEI/10.13039/501100011033/, and by ``ERDF A way of making Europe'', European Union. G.E.R. also acknowledges support from PIP 0554 (CONICET). F.G. acknowledges support from PIBAA 1275 and PIP 0113 (CONICET). S.d.P. acknowledges support from ERC Advanced Grant 789410.

\end{acknowledgements}

\bibliographystyle{aa}
\bibliography{biblio,morerefs,refs}

\begin{appendix}

\section{Marginal potential candidates}\label{app:marginal}

For the sake of completeness, in Table~\ref{Table:MoreFRB} we report the marginal FRB candidates found in this survey. Where the search recovered several closely spaced trials of the same event at neighbouring DMs, only the trial with the highest FETCH $p$-score is listed. The final downselection was not based solely on the FETCH $p$-score: all triggers were also inspected visually, and candidates whose diagnostic plots suggested contamination by radio-frequency interference were discarded, even when their $p$-scores were high.

\begin{table*}
\centering
\caption{Marginal FRB candidates identified in the Sculptor and Phoenix region observations.}
\label{Table:MoreFRB}
\begin{tabular}{ccccccc}
\hline\hline
Field & Date & $t_\mathrm{start}$ (MJD) & $t_\mathrm{cand}$ (s) & DM ($\dmunits$) & S/N & $p$-score \\
\hline
A4013 & 2025-10-18 & 60966.115300925929 & 1073.0615600 & 317.00 &  8.41 & 0.9822453 \\
A4013 & 2025-10-15 & 60958.148634259262 & 11126.0998040 & 364.00 &  9.23 & 0.9970598 \\
A2870 & 2025-10-07 & 60955.111828703702 & 1102.9774340 & 470.00 &  8.40 & 0.8548032 \\
A2870 & 2025-09-16 & 60934.169467592590 &  1160.6419050 & 488.00 &  8.90 & 0.6974987 \\
A2870 & 2025-06-20 & 60846.384050925924 &    17.0675400 & 240.00 & 12.47 & 0.5401518 \\
\hline
\end{tabular}
\tablefoot{Dates are UTC dates of the start of the observation. Where several DM trials of the same event were recovered, only the trial with the highest FETCH $p$-score is listed.}
\end{table*}

\begin{figure}
    \centering
    \includegraphics[width=\hsize]{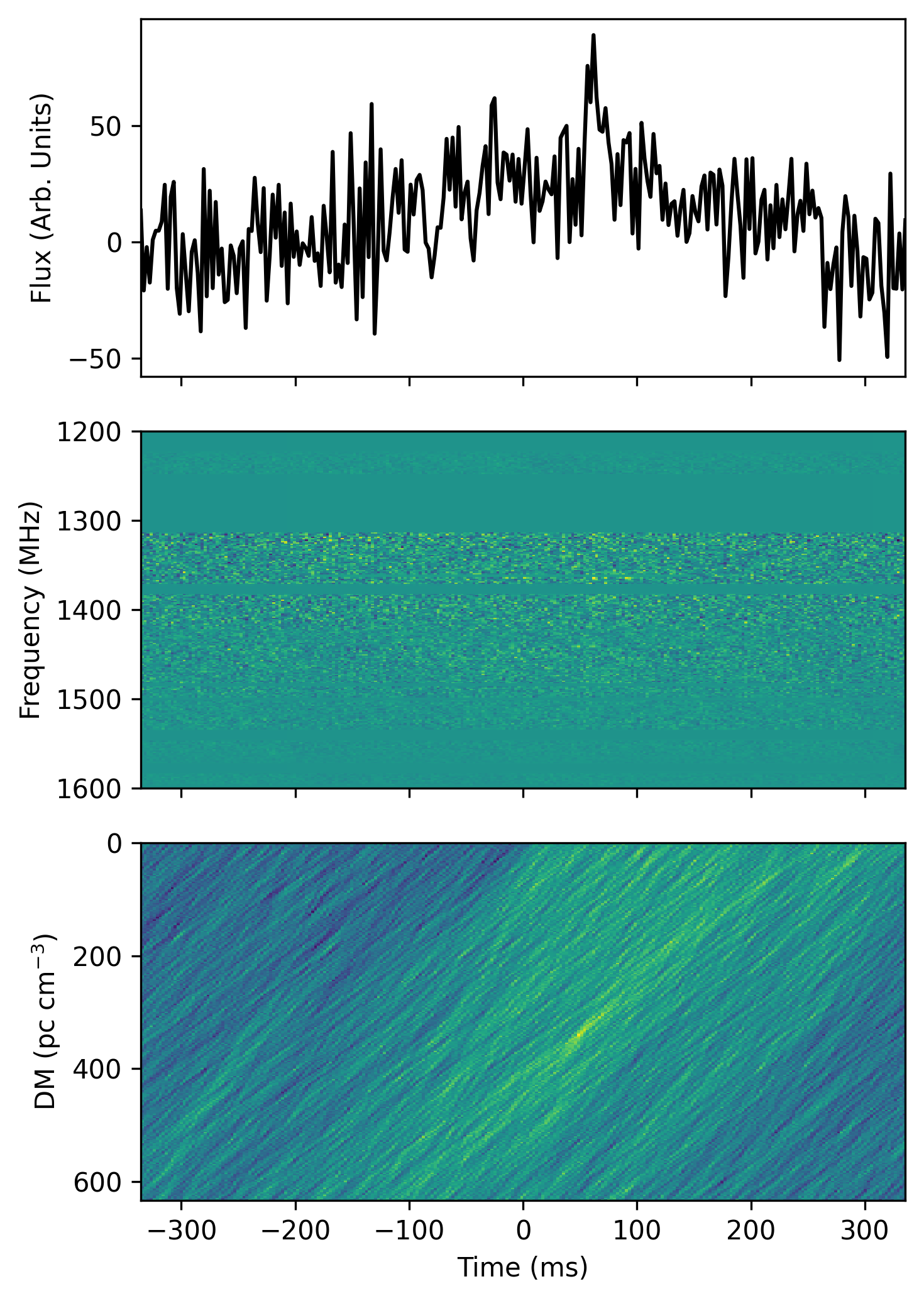}
    \caption{Marginal FRB candidate at MJD 60966.12772061 in the direction of the Sculptor cluster A4013.}
    \label{fig:FRB2}
\end{figure}

In the observations of the Sculptor Abell cluster A4013 (declination $\delta=-35\degr 16\arcmin 00\arcsec$, right ascension $\alpha=$ 23h 31m 54s), we identified, in the observation of 18 October 2025 starting at MJD 60966.115300925929, a candidate at $t=1073.0615600$~s, with a dispersion measure $\mathrm{DM}=317~\dmunits$, a signal-to-noise ratio $\mathrm{S/N}=8.41$, and a FETCH-assigned FRB probability of $p=0.98$. Figure~\ref{fig:FRB2} shows this FRB candidate.

\section{Details of the validation tests}\label{app:Tables}

In order to test our ability to identify FRBs in radio-astronomical data, we first performed a search on Parkes (Australia) data containing known FRB detections \citep{Keane:2014lja}, available at \url{http://researchdata.ands.org.au/fast-radio-bursts-parkes/468266}. This archival data set comprises seven FRBs (FRB~010125, FRB~010621, FRB~010724, FRB~110220, FRB~110626, FRB~110703, and FRB~120127), with full-resolution SIGPROC filterbank files for all 13 beams of each pointing; the data for FRB~140514 are available separately\footnote{\url{https://researchdata.edu.au/fast-radio-burst-frb-140514/468269}}. In particular, we focused on the detection of FRB~110703, and we also injected a synthetic signal with \texttt{FRB-faker} that was successfully recovered by our pipeline.
The filterbank data were processed and searched for bursts using the Heimdall single-pulse software package\footnote{\url{http://sourceforge.net/projects/heimdall-astro/}} and, independently, with \texttt{PRESTO}\footnote{\url{https://github.com/scottransom/presto}}; the resulting candidates were then passed to FETCH\footnote{\url{https://github.com/devanshkv/fetch}} to assess whether they were genuine FRBs. We successfully recovered the FRBs with the correct arrival times and dispersion measures. We also found in our limited study that \texttt{PRESTO} led to fewer false-positive
detections than Heimdall for the IAR observations, hence we adopted it by default in our searches.

As an additional blind test, in August 2024 we injected a synthetic single pulse (DM $=874~\dmunits$, $\mathrm{S/N}=18$, width 80~ms, at $t=234$~s) with \texttt{FRB-faker} into an IAR filterbank observation. The file was then independently analysed by three members of our team, all of whom recovered the injected burst with the pipeline described in Sect.~\ref{sec:Techniques}.

Our next test made use of actual observations from the IAR. On 2024-09-23 we carried out five observations of 5 minutes (300~s) each, pointing antenna A2 at the zenith, and injected one synthetic FRB into each of the observations using \texttt{FRB-faker}. All injected bursts shared the same morphology---a pulse profile consisting of two subcomponents with widths (FWHM) of 2.8 and 4.0~ms, separated by $\sim$4~ms, each modelled as a Gaussian in the frequency domain (centred at 1463 and 1419~MHz with FWHMs of 180 and 250~MHz, respectively) with a spectral tilt of 8---but each had a different DM, injection time, and S/N, as listed in Table~\ref{tab:FRBFaker}.
\begin{table}
  \centering
  \caption{Synthetic FRBs injected into test observations 1--5.}
   \label{tab:FRBFaker}
   \begin{tabular}{cccc}
     \hline\hline
     Observation & DM ($\dmunits$) & Time (s) & S/N \\
    \hline
     \#1 & 560 & 160 & 28 \\
     \#2 & 280 & 256 & 10 \\
     \#3 & 410 &  86 & 45 \\
     \#4 & 120 & 227 & 32 \\
     \#5 & 356 & 136 & 17 \\
   \hline
   \end{tabular}
 \end{table}
We were able to successfully recover all of these injected FRBs using the techniques described above.
For the system parameters adopted in Sect.~\ref{sec:Discussion}, the $8\sigma$ detection threshold of our pipeline corresponds to a minimum detectable fluence of $\mathcal{F}_\mathrm{min} \approx 7\,(W/\mathrm{ms})^{1/2}$~Jy\,ms (equivalently, a minimum peak flux density of $\approx 7\,(W/\mathrm{ms})^{-1/2}$~Jy) for a burst of width $W$ and the nominal 264~MHz of clean bandwidth, degrading to $\approx 9\,(W/\mathrm{ms})^{1/2}$~Jy\,ms in observations for which per-observation flagging leaves only $\simeq 170$~MHz, as was the case for the detection reported here. To these values must be added a $\sim$16\% systematic uncertainty inherited from $T_\mathrm{sys}$. Together they define the fluence completeness of the survey.

\end{appendix}

\end{document}